# Modelling the Effectiveness of Curriculum in Educational Systems Using Bayesian Networks

Ahmad A. Kardan
Department of Computer Engineering
Amirkabir University of Technology
424, Hafez, Tehran, Iran

Omid R. B. Speily
Department of Computer Engineering
Amirkabir University of Technology
424, Hafez, Tehran, Iran

Yosra Bahrani
Department of Computer Engineering
Amirkabir University of Technology
424, Hafez, Tehran, Iran

*Abstract*—**in recent years, online education has been considered as one of the most widely used IT services. Researchers in this field face many challenges in the realm of Electronic learning services. Nowadays, many researchers in the field of learning and eLearning study curriculum planning, considering its complexity and the various numbers of effective parameters.**
**The success of a curriculum is a multifaceted issue which needs analytical modelling for precise simulations of the different learning scenarios. In this paper, parameters involved in the learning process will be identified and a curriculum will be propounded. Furthermore, a Curriculum model will be proposed using the behavior of the user, based on the logs of the server. This model will estimate the success rate of the users while taking courses. Authentic Bayesian networks have been used for modelling. In order to evaluate the proposed model, the data of three consecutive semesters of 117 MS IT Students of E-Learning Center of Amirkabir University of Technology has been used. The assessment clarifies the effects of various parameters on the success of curriculum planning.**

*Keywords-* *Online education ;Electronic learning services; Curriculum model; curriculum planning; Bayesian networks*

## I. INTRODUCTION

Nowadays, online education has made extreme advances alongside other IT services. With regards to its user friendliness, inexpensiveness, and providing a variety of different services, online education could turn into one of the leading IT services. In spite of its increasing advances, this field has many challenges. Providing new services to improve the quality of online education has been of utmost significance in this field of research for scholars and researchers and providing online education in the worldwide web allows numerous users with different attributes to use this service. In addition to the many challenges that may occur due to the variety and number of users, systems can provide innovative services for their users while collecting crucial and valuable data about users' interactions and analyzing this data. One of the new services of the online education system is to give a simulated system of a personalized curriculum. That is, by considering the users educational state and the given courses, the system can evaluate the user's curriculum from various perspectives. This evaluation may assist users in taking courses and curriculum planning in the form of an intelligent DSS or expert system.

This type of system uses different statistical methods, data-mining and investigating the state of users and their courses to identify different parameters in the success rate of users.

In this paper, an attempt has been made to investigate numerous graduate students of the E-Learning Center of Amirkabir University of Technology in order to identify these parameters in Information Technology field. The Bayesian network is identified using effective parameters including these variable forms. This network determines the possibility of investigating the success of curriculum planning and specifies the effectiveness rate of variables based on data set in this research. In this study, we used version 3 of SamIam tools of California (Los Angeles) University to create and analyze the Bayesian network.

The paper is organized as follows. First in section 2, a review of previous works in this field is presented. Then in section 3, in addition to giving suggested models for curriculum planning, we describe parameters and methods for making Bayesian networks. Finally, in sections 4 and 5, by evaluating the suggested model, a summary of this research is given.

## II. RELATED WORK

Many of the research studies carried out thus far in this realm have not been focused on the engineering basis of the issue. In this section, we review some relevant works carried out. Since the Student Scientific Apprenticeship Model has recently been very well validated, the educational model of secondary students is targeted on the basis of the Student

Scientist Apprenticeship. The outcomes of this model are an increasing interest in scientific professions, tendency to continue with higher education, and positive outcomes resulting from on-job training experiences[1]. Participating in training programs keep students in line with science and research. The resulting outcomes of this model rely on the confidence for carrying out organizational operations, intellectual development and understanding the scientific context. In fact, the output of this model is interpreted and divided into some categories as follows: professional goals, knowledge of the scientific context, confidence, self-efficacy, intellectual development and skills. In [2], two points are considered: (1) knowing knowledge that cannot be learned from other courses, and (2) providing good opportunities for organizations to choose students proportionate to the condition of organizational work. The aim of this study is to introduce a curriculum to decrease gaps between theory and practice. In this course, a teaching method is presented according to real experiences resulting from collaboration with schools. In fact, organizational collaboration courses have emphasis on transforming from theory to practice. The integrated educational collaboration model is represented in order to find real requirements of organizations and new technology for developing students' skills in practice, ability of creativity, collaboration skills and other general skills.

In [3], the aim of improving educational outcomes is based on individual skills and preparing students for organizational work. Four curriculum models are analyzed on this basis: the educational model of Germany's dual system, the model of the Australian TAFE curriculum, the model of DACUM and the international model of the MES curriculum. In fact, analyzing these models is to remove problems related to lack of personnel with adequate skills in the organizations, forming technical and vocational curriculum, and optimizing the technical structure of organizations.

III. PROPOSED METHOD

In this section we will propose a method for modeling the curriculum. Prior to giving details on the proposed model, the basics and concepts of Bayesian networks and their role will be presented.

A. *Bayesian networks and their roles in online education*

Considering that curriculum modeling of learners has an uncertainty structure, a structure is required in which this uncertainty is very well considered. The Bayesian network is a complete model for unstructured data to show learner's modeling and it can be a model for variables and their relationships [4], [5]. Here we examine the principles of the Bayesian network and infer related methods, as taken from [8].

B. *Introducing Bayesian networks*

The Bayesian network is a graph without cycles and directions in which nodes present "concepts" and edges are relations of the cause and effect between nodes. This graph can be applied for evaluating the probability of carrying out each activity by learners or evaluating the probability of learning knowledge by them. Bayesian reasoning provides a method based on the probability for inferring. This method is based on probability distribution and optimizing decisions can be made by reasoning on this probability with observed data. This method is very important in machine learning because it provides the quantity solution for weighting evidence that supports different assumptions. Bayesian reasoning provides a direct method for supervised algorithms, providing a framework for algorithms dealing with probability [8], [9].

C. *Bayesian Theory*

In most cases, finding the best assumption in the assumed space H along with having training data D is necessary. One method to state the best assumption is that we follow the most probabilistic assumption, in addition to having data D and the primary knowledge about probabilities of assumption H. The Bayes' rule provides a direct method for evaluating this probability.

Some symbolization is necessary to define Bayes' rule. Before considering some training data, P (h) is used to state the primary probability that p assumption is correct. Generally P (h) is called prior probability and indicates the previous knowledge dependent to the chances of a correct assumption h. If we do not have any knowledge of the assumptions, we can appropriate one equal probability to the entire assumption space H. Similarly, P(D) is used to indicate the prior assumption that data D is observed (in other words, the probability of observing D when there is no knowledge about the correctness of the assumptions). Also P (D|h) is used to indicate the probability of D in a world in which assumption h holds. In learning machine P (h|D) is the desired subject, that is, the probability of a correct assumption of h in the case of observing training data D. P (h|D) is called the posterior probability of h, so it indicates our confidence in the assumption h after observing data D. Bayes' rule is a principle part in learning Bayes, because it provides a method for evaluating the posterior probability P(H|D) from prior probability along with P(D) and P(D|h) [5],[10],[11].

$$P(h|D) = \frac{P(D|h)P(h)}{P(D)} \quad (1)$$

As expected, it can be observed that P (h|D) is increased by an increase of P(h) as well as P(D|h). In turn, it seems logical that P(D|h) is reduced by an increase of P(D), because when increasing the probability of P(D) occurring that is independent of h, there is less evidence in D for supporting h. In many learning scenarios, learners consider a series of H assumptions and are interested in finding

assumptive $h \in H$ that is most probabilistic (or at least is one of the most probabilistic if some exist). Any assumption that has these characteristics is named the Maximum a posteriori assumption (MAP) [6], [7], [8]. The MAP assumption can be obtained by using Bayes' rule for evaluating posterior probability. In exact terms, $h_{MAP}$ is an assumption that:

$$h_{MAP} = \arg\max_{h \in H} P(h \mid D)$$
$$= \arg\max_{h \in H} \frac{P(D \mid h)P(h)}{P(D)} \quad (2)$$
$$= \arg\max_{h \in H} P(D \mid h)P(h)$$

Notice that in the last step above, P(D) is removed, because its evaluation is independent of h and is always a constant number. In some cases it is assumed that all assumptions of $h \in H$ have equal probability of occurrence $\forall h_i, h_j \in H; P(h_i) = P(h_j)$. In this case, other simplifications can be done in equation 2 [9], [7]. On the other hand, we can consider the assumption that maximizes P (D|h) (Maximum likelihood).

$$h_{ML} = \arg\max_{h \in H} P(D \mid h) \quad (3)$$

Here we consider data D as educational sampling for the target function and H is considered a series of target functions. But in fact, the Bayes' rule is more general than this issue. That is, it can be used in a similar way for a series of any kind of assumptions that are independent two by two, with a total probability of one.

*D. Simple Bayesian classification*

The Simple Bayesian learner's method is one of the more applicable Bayesian training methods that are often called simple Bayesian classification. In some fields it is shown that its efficiency is comparable with the efficiency of methods such as the neural network and decision tree. This section introduces a method of simple Bayesian classification. The next section shows an application of this method in a practical example.

Simple Bayesian classification is applied for instances in which every sample X is selected by a series of adjective amount and target function f(x) from a series like v. The series of training data and output of the target function or class that the new sample belongs to is targeted. The Bayesian method for classifying a new sample is to identify the most probabilistic class or $v_{MAP}$ target amount by having an adjective amount <a1, a2,…, an> describing the new sample.

$$v_{MAP} = \arg\max_{v_j \in V} P(v_j \mid a_1, a_2, ..., a_n) \quad (4)$$

By using Bayes' rule we can rewrite equation 4 as:

$$v_{MAP} = \arg\max_{v_j \in V} \frac{P(a_1, a_2, ..., a_n \mid v_j) P(v_j)}{P(a_1, a_2, ..., a_n)} \quad (5)$$
$$= \arg\max_{v_j \in V} P(a_1, a_2, ..., a_n \mid v_j) P(v_j)$$

Now we attempt to estimate the two statements of equation (3) using training data. Evaluating training data is easy if we know the rate of $v_i$ repetition. However, evaluating different sentences P $(a_1, a_2, ..., a_n \mid v_j)$ is not acceptable unless we have a lot of training data. The problem exists where the number of sentences equals the number of samples multiplied by the amount of the target function. Thus, we must observe every sample several times until we obtain an appropriate estimate [11].

The assumption of a simple Bayesian classification method is based on this simplification that adjective amounts having amounts of the target function are conditionally independent of each other. On the other hand, this assumption indicates that in the case of observing the output of the target function, the probability of observing adjectives a1, a2,… is equal to multiplying the probability of each adjective separately. If this is replaced by equation (4), it will result in a simple Bayesian classification method.

$$v_{NB} = \arg\max_{v_j \in V} P(v_j) \prod_i P(a_i \mid v_j) \quad (6)$$

Where $v_{NB}$ is the output of the simple Bayesian classification method for the target function. Notice that the number of sentences of $P(a_i \mid v_j)$ that must be evaluated in this method is equal to the number of adjectives multiplied by the number of output category for the target function, the amount of which is much less than the number of sentences of P $(a_1, a_2, ..., a_n \mid v_j)$. The result is that simple Bayesian training attempts to estimate different amounts of $P(a_i \mid v_j)$ and $P(v_j)$ by using their repetition rate in the training data. This series of estimations correspond to the learned assumptions. Then this assumption is used to classify the new samples, which are carried out by using equation (4). Whenever the independent assumption of conditioning the simple Bayesian classification method is satisfied, the simple Bayesian class will equal the MAP class [11], [12].

*E. Curriculum Modelling*

For curriculum modelling, we initially determine the effective parameters on the success or lack of success of a selected curriculum. The selected parameters for curriculum modelling in the Ms are shown in table 1. Also, the grade of the current term, rate of satisfaction and chance of receiving a letter of recommendation are other effective parameters in an effective curriculum.

In the terminology of Bayesian networks, these parameters are considered as variables. Three steps should be derived for modelling using the Bayesian network [12].
1) A series of relevant parameters along with its amount should be derived.
2) The network structure should be in the form of a graph without cycles and with nodes of variables.
3) For every network variable, a CPD (Conditionally Probability Distribution) is determined.

These previous three steps are carried out for curriculum modelling using Bayesian networks, respectively. The effective parameters are determined by expert opinion and available data. The amount of these variables is specified in Figure 2.

In the second step, the Bayesian network structure is formed with the graph without cycles. The Bayesian network structure is shown in Table 1. It should be noted that the direction of edge between two variables determines the effectiveness of the two variables. For example, variables of the total average are effective on the current term's GPA (grade point average). Thus, the current term's average is effective on the student's satisfaction rate of the curriculum.

In the third step, the amount of CPD for the variables of the network is determined. A sample of CPD associated with grade is observed in Table 2. The grade has three amounts with specified probabilities according to this table.

TABLE 1: PARAMETERS EFFECTING CURRICULUM

| No | Parameters |
|----|------------|
| 1 | Average Grade |
| 2 | Number of selected course |
| 3 | State |
| 4 | Acivity |
| 5 | Publications |
| 6 | Research BackGround |
| 7 | Satisfaction |

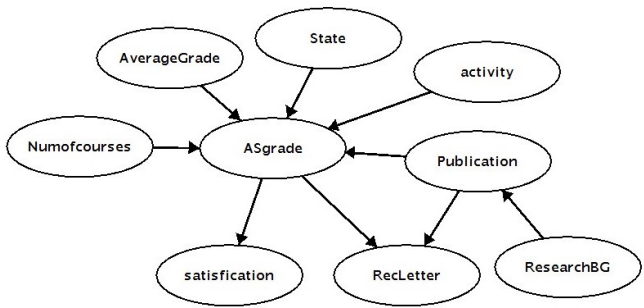

Figure 1: Bayesian Network

TABLE2: SAMPLE OF CDP RELATED TO TOTAL AVERAGE

| Ave Grade | Prop |
|-----------|------|
| A | 0.41 |
| B | 0.3 |
| C | 0.29 |

The amount of P (AG, S, A, NumC, RBG, Pub, G, RecL, satisfaction) based on the chain rule of Bayesian networks and on the basis of network structure for the curriculum is equal to:

$$P(AG,S,A,NumC,RBG,Pub,G,RecL,Satisficatin) \\ = P(AG)P(S)P(A)P(NumC)P(RBG)P(Pub|R) \\ P(G|AG,S,A,NumC,RBG,Pub)P(RecL|G,Pub) \\ P(Satisfaction) \quad (7)$$

in which AG is last term's average grade, S is the state of the student, A as activity, NumC is the number of lessons, RBG is the research background, Pub is published papers, G is the current term grade, RecL is recommendation letter and satisfaction is rate of satisfaction.

For each of the variables in the relation above, CPD should be determined with regard to its own data set.

## IV. THE RESULTS OF MODELLING

In this paper, we used version 3 of the SamIam tool for curriculum modelling and the Bayesian network. Figure 1 and 2 show a part of the Bayesian networks designed by this tool. The data set is considered for 117 M.S. students of IT from the E-Learning Center of Amirkabir University of Technology in three continuous terms. All amounts of CPDs are determined with regard to expert opinion and field operation on the data sets. Results of modelling determine the rate of direct and indirect effectiveness of the variables.

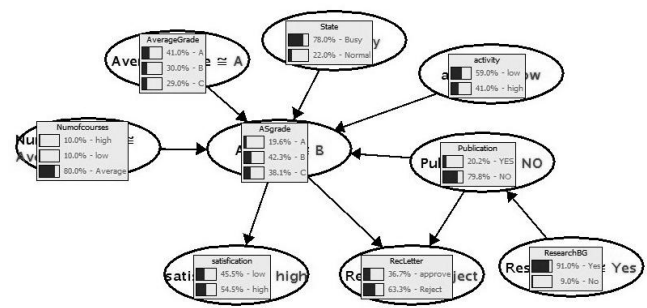

Figure 2: A shear of Bayesian networks designed by SamIam

As shown in Figure 3, according to the statistical data set, the greatest effect in recommendation letter approval is given to the current term's GPA and published papers with effectiveness of 1.95 and 0.93, respectively. When we investigated the data set for not receiving a recommendation letter (as shown in figure 4), we have similar variables as before, but with different effectiveness. That is, the effectiveness rate of the current term's GPA and published papers with amounts 1.05 and 0.93 respectively, obtain the highest amount once more.

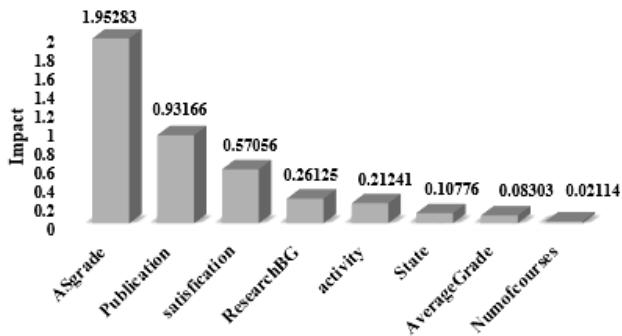

Figure 3: Parameters that affect approval of recommendation letter and its impact level

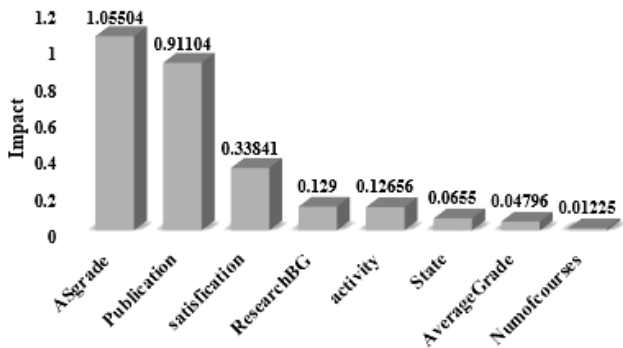

Figure 4: Parameters that affect the rejection of recommendation letter and its impact level

The student's satisfaction of the curriculum is shown in Figures 5 and 6. Both factors, the current term's GPA and receiving recommendation letter, are recognized as the most effective factors in the rate of satisfaction. However, the effectiveness factor of the current term's GPA has had a more significant effect on student's overall satisfaction.

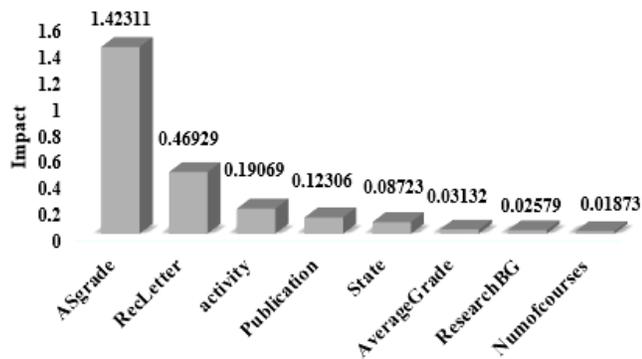

Figure 5: Parameters that affect the low satisfaction of students and its impact level

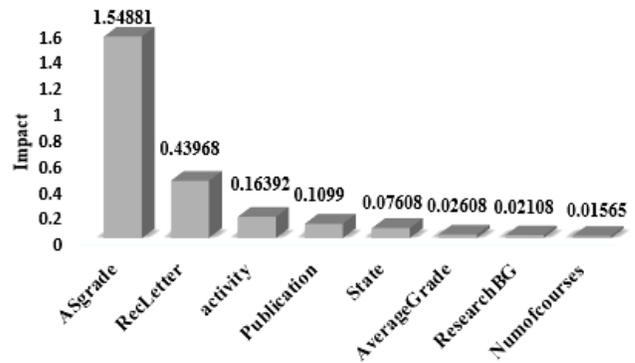

Figure 6: Parameters that affect the high satisfaction of students and its impact level

In this part of analysing the curriculum model we deal with investigating the condition of the current term GPA in three amounts of A, B and C. As shown. In Fig. 7, the current term average of A has the greatest effect on student satisfaction. The rate of this satisfaction is estimated at about 1.78. After rate of satisfaction, the average A has the most effectiveness on the variables such as recommendation letter and activity with amounts of 1.32 and 0.63, respectively.

The current term average of B has quite a different effectiveness that it is shown in Fig.8. It should be considered that this effectiveness can be negative or positive. In this state, the most effectiveness belongs to the total GPA. The rate of effectiveness of this variable is about 0.43. This effectiveness can be justified in this manner that the overall average has a close link with the current term average. In other words, the total average is usually repeated in the current term.

In this manner, results of the effective factors on the current term's GPA are shown in Fig. 9. The most (though negative) effectiveness on recommendation letter will be 1.68. In addition to effect on recommendation letter, both parameters of student satisfaction and activity, with amounts of 1.2 and 0.41, have the most cross-effect, respectively.

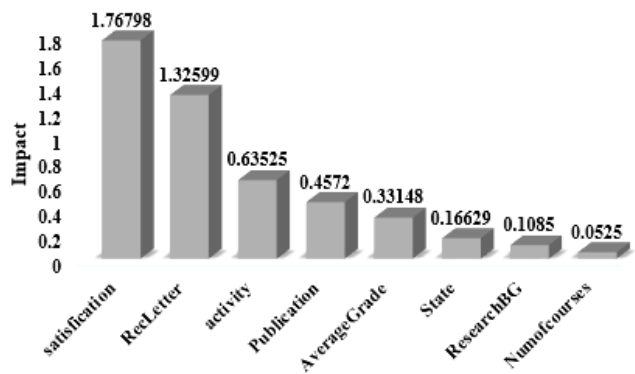

Figure 7: Parameters that affect the Average A of students and its impact level

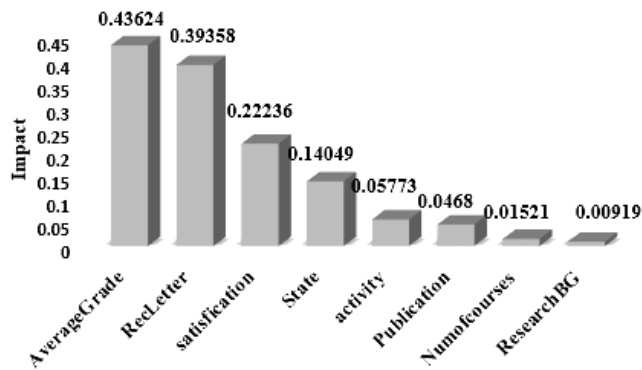

Figure 8: Parameters that affect the Average B of students and its impact level

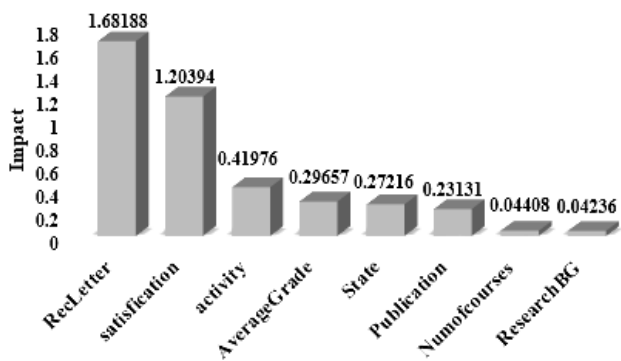

Figure 9: Parameters that affect the Average C of students and its impact level

## V. CONCLUSION

In this paper we deal with the subject of curriculum planning in terms of course selection and its effective factors. Thus, a model of curriculum planning and its effective parameters should be designed. In this study, the method of Bayesian networks is used for curriculum modeling. The effective parameters are initially derived, and then the Bayesian network structure is formed while depicting a graph without directional cycle. While investigating data associated with 117 M.S. students of E-Learning Center of Amirkabir University of Technology, CPDs are initialized. This study is designed and created by the use of version 3 of the SamIam tool of the Bayesian network. While analyzing the cross-effect on several variables, interesting results are achieved for curriculum management.